\documentclass{iosart2c}

\usepackage[T1]{fontenc}
\usepackage{times}%
\usepackage{subcaption}
\usepackage{amsmath}
\usepackage{dcolumn}
\usepackage{graphicx}

\newcolumntype{d}[1]{D{.}{.}{#1}}

\firstpage{} \lastpage{} \volume{} \pubyear{2020}

\begin{document}
\begin{frontmatter}                           

%
\title{Radiomic feature selection for lung cancer classifiers}

\runningtitle{Radiomic feature selection for lung cancer classifiers}

\author[A]{ \snm{Hina Shakir}\thanks{Corresponding author. E-mail: hinashakir.bukc@bahria.edu.pk.}},
\author[A]{\snm{Haroon Rasheed}}
and
\author[B]{\snm{Tariq Mairaj Rasool Khan}}
\runningauthor{Hina Shakir et al.}
\address[A]{Department of Electrical Engineering, Bahria University, Karachi, Pakistan \\
E-mail: \{hinashakir.bukc,haroonrasheed.bukc\}@bahria.edu.pk}
\address[B]{Department of Electrical and Power Engineering, PNEC, National University of Science and Technology, Pakistan \\
E-mail: khan.tariq@pnec.nust.edu.pk}

\begin{abstract}
Machine learning methods with quantitative imaging features integration have recently gained a lot of attention for lung nodule classification. However, there is a dearth of studies in the literature on effective features ranking methods for classification purpose. Moreover, optimal number of features required for the classification task also needs to be evaluated. In this study, we investigate the impact of supervised and unsupervised feature selection techniques on machine learning methods for nodule classification in Computed Tomography (CT) images. The research work explores the classification performance of Naive Bayes and Support Vector Machine(SVM) when trained with 2, 4, 8, 12, 16 and 20 highly ranked features from supervised and unsupervised ranking approaches. The best classification results were achieved using SVM trained with 8 radiomic features selected from supervised feature ranking methods and the accuracy was 100\%. The study further revealed that very good nodule classification can be achieved by training any of the SVM or Naive Bayes with a fewer radiomic features. A periodic increment in the number of radiomic features from 2 to 20 did not improve the classification results whether the selection was made using supervised or unsupervised ranking approaches.
\end{abstract}


\end{frontmatter}

\section{Introduction}
Lung cancer has remained a major cause of deaths among cancer patients for the last few decades \cite{Siegel}. Besides better screening facilities, there is an emerging interest in computer-aided diagnosis (CAD) systems to improve the detection rate of cancer. The CAD systems offer good potential in cancer diagnosis and can assist radiologists and clinicians in decisions making before the actual symptoms manifest \cite{Ayman}. Thus, many lives can be saved by cancer detection on medical images.
Currently, nodule classification through machines learning classifier involves extracting quantitative imaging features of a lung nodule and employing machine learning techniques to train a classification model with discriminative features towards cancer. The quantitative imaging features are commonly termed as radiomic features and have shown a strong correlation with lung cancer diagnosis as well as prognosis \cite{Gillies}. The extracted radiomic features from lung nodules are fed into a classifier model which classifies the nodule as benign or malignant.

For machine learning based classification, Chen et al. in \cite{Chen} proposed a radiomic signature of four law features including LSLmin, SLLenergy, SSLskewness and EELuniformity and used Support Vector Machine as a nodule classifier. The feature selection was performed using Wilcoxon rank sum test and Sequential Forward Selection (SFS). The proposed  signature classified the nodules with an accuracy of 84\%, sensitivity of 92.85\% and the specificity of 72.73\% respectively. Ma et al. \cite{Jingchen} extracted 583 features from the lung nodule CT images and classified the nodules using random forest approach. The proposed approach achieved an accuracy of 82.7\%. Tu et al. in \cite{Tu} investigated 238 features out of 374 features to be useful to differentiate between benign and malignant nodules. The chosen features were employed to train a set of high performance machine learning classifiers for early detected cancer diagnosis in thin-section CT. In the study conducted, logistic classifier showed better results than the other classifiers and achieved an accuracy of 79\%. Kadir and Gleeson \cite{Kadir} presented a study that if sufficient training data is present then convolutional network with deep learning can classify the nodules with area under curve in the ranges of 0.90. Feature selection was an automatic process where 15 features were automatically selected from 23 features possibilities. 

\par Choi et al. \cite{Choi} showed a radiomics based classification model for lung nodules using SVM LASSO classifier trained on 2 radiomic features with 5 fold and 2 fold Cross-validations (CVs) with accuracy of 84.1\% and 81.6\% respectively. The features were selected applying univariate analysis with Wilcoxon rank-sum test and Area under the curve (AUC) to evaluate the significance of each feature. The authors compared their presented model with the Lung CT Screening Reporting and Data System (Lung-RADS) and showed through quantitatively analysis the superior performance of their approach. Wu et al. \cite{Wu} selected the diagnostic features using minimum correlation between the features followed by univariate analysis. The authors evaluated the classification performance of 53 discriminative features using Random Forests, Naive Bayes algorithm and K-nearest neighbors.The experiments revealed that Naive Bayes outperformed the other methods and classified the nodules using 5 features with an AUC of 0.72. Tao et. al \cite{Sun} showed that application of textural features of a lung nodule transformed using curvelets to a SVM increases the classification performance of early stage cancer. The area under the curve achieved was 0.949 and the obtained accuracy for the unbalanced and balanced data was 80\% and 90\% respectively.

\subsection{Contribution of the proposed work}
The radiomic features selection in the above mentioned machine learning classification models was either performed using feature reduction techniques or with a fewer features chosen due to their discriminative power towards cancer using a variance test. \par After the application of feature reduction techniques, features selection methods can help in finding the capable features which can differentiate between benign and malignant tumors.  Therefore, it is crucial to investigate the role of feature selection techniques and to evaluate the usefulness of supervised and unsupervised selection methods in context of tumor classification. While there is a dearth of comparative studies on feature selection techniques, the optimal number of features required to achieve better performance of machine learning classifiers also needs to be researched. 

In this paper, we carry out a performance analysis of supervised and unsupervised feature selection techniques for two machine learning  methods including Support Vector Machine(SVM) and Naive Bayes to classify a lung nodule as benign or malignant. First, a group of 20 discriminative features was obtained using averaged scores of supervised as well as unsupervised ranking methods. Then, the number of radiomic features applied to Naive Bayes and SVM were periodically increased from 2 to 20 for classifier training. The trained machines were employed for the classification of 50 malignant and 30 benign nodules. The features chosen using supervised feature selection algorithms outperformed the ones selected through unsupervised ranking algorithms when classification results were evaluated and compared in terms of accuracy, specificity and sensitivity. The experiments also showed that increasing the number of radiomic features periodically from 2 to 20 obtained using supervised ranking methods did not make any positive impact on the classification results. 

\par The remainder of the paper is organized as follows. Section 2 reviews the classification models and feature selection methods used in this research work. Section 3 outlines the  methodology adopted for radiomic feature selection and training of machine learning classifier. The results of nodule classification followed by a discussion are presented in section 4 and 5 respectively. Finally, the conclusion is drawn in section 6.

\section{Classification Models and Feature selection methods}
\subsection{Support Vector Machine versus Naive Bayes as machine learning methods}
The proposed radiomic feature analysis is performed over the two popular and the most frequently used machine learning classifiers including Support Vector Machine and Naive Bayes algorithm.

A Support Vector Machine(SVM) is a supervised learning model which is used for regression and classification. SVM performs classification by defining a separating hyper-plane between distinguishing features. In practice, a SVM is trained with the labeled data which is also called supervised learning where the algorithm generates an optimal hyper-plane to categorize the test data into the provided labels \cite{Shi}. For the test data with two labels, this hyper-plane is a line which divides a plane in two parts where the labels of each class are on the either side. A good classification is achieved when the generated hyper-plane has the largest distance to the nearest training-data point of any class which in turn results in lower generalization error of the classifier.

On the other hand, a Naive Bayes classifier is a supervised machine learning model which uses Naive Bayes algorithm for the classification purpose. The algorithm computes the joint distribution p(a,b) of the extracted features $a$ and the class labels $b$ given by $p(a \mid b)p(b)$ and then learns the parameters of model by maximizing its likelihood function \cite{Shi}. The relationship of the labels and the features learnt through the above steps is stable and the classification results do not change in general for the noisy data. 

\subsection{Supervised and unsupervised feature selection algorithms}

The most discriminative features towards cancer were obtained by ranking the features through a group of supervised as well as unsupervised ranking algorithms. Supervised feature selection takes into account the feature class label(malignant versus non-malignant) along with its associated feature for ranking purpose. Features showing dissimilar values for different classes are ranked higher than the ones which exhibit similar values for different classes \cite{Shakir1}.

One the other hand, in unsupervised feature selection algorithms, the inherent traits of features are learned and patters are inferred to discriminate between similar and dissimilar features. For this purpose, statistical measures are used and class labels do not contribute in ranking or selection. 
\section{Methodology}
\label{usage}

 The work flow of the proposed experimental setup is shown in Figure 1. The lung nodules from acquired CT data sets were segmented and 3-D radiomic features for every segmented nodule were computed. Supervised and unsupervised feature selection algorithms were employed for feature ranking, and highly ranked features were used to train machine learning classifiers for nodule classification. The test lung nodules were then classified as benign or malignant using trained model. 
The experimental CT data sets comprised of 215 malignant nodules acquired from Lung1 database \cite{Hugo} and 35 and 29 benign nodules were accessed from LIDC \cite{Armato} and LUNGx \cite{Armato1} databases respectively. The acquired data sets were divided into training and test cohorts. Description of the lung nodules databases and the average nodule sizes is given in Table 1.

\begin{figure}[!tbp]
{\includegraphics[width=8.08cm,height=11.89cm]{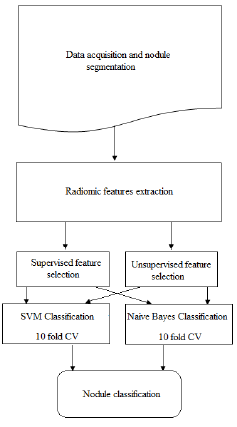}}
\caption{Work flow of the proposed experimental setup}
\label{fig}
\end{figure}

\subsection{Lung nodules segmentation and radiomic feature extraction}
The nodules segmentation of lung1 data sets was performed using the manual segmentation information provided with the database. The Grow Cut algorithm from the Slicer Platform \cite{Fedorov} was employed to segment the CT volumes of LUNGx and LIDC datasets. Segmentations performed by Grow Cut algorithm have proven to be highly consistent with the manual segmentations \cite{Velazquez}. A 2-D and 3-D axial slice of a test lung nodule segmented from Lung1 CT data set is shown in Figure 2. 
\par Following the nodule segmentation, a total of 105 3 -D radiomic features were computed from the segmented lung nodules using PyRadiomics module \cite{Griethuysene} of Slicer platform. The computed features belonged to the features classes of shape (n= 13), first order statistics(n=19), Gray level Difference Method (GLDM)(n=14), Gray-Level Co-Occurrence Matrix (GLCM)(n=23), Gray Level Size Zone Matrix (GLSZM) (n=16), Gray Level Run Length Matrix (GLRLM) (n=16) and Neighborhood Gray-Tone Difference Matrix (NGTDM) (n=5). The shape class encompasses features describing the shape of the nodule, whereas the First order statistics features describe the texture of the nodule. The other five classes including GLDM, GLCM, GLSZM, GLRLM and NGTDM measures the gray level intensity within the nodule using second order statistics and quantitative methods including size zone, run length, differences and co-occurrence.

\subsection{Pre-processing step and Feature ranking using supervised and unsupervised feature selection algorithms}
In order to select the distinguishing features towards cancer, one-way ANOVA test for 5\% significance level was performed on the computed radiomic features of 200 training data sets. Based on the test results, 52 features were found to be discriminative towards lung cancer. The distribution of the selected features and the extracted features with respect to their feature classes is shown in Figure 3. The selected features constitute of 10 features from Shape class, 8 features from GLDM class, 12 features from GLCM class, 8 features from First order class, 7 features from GLRLM, 6 features from GLSZM and 1 feature from NGTDM class.

Three efficient supervised feature selection algorithms \cite{Kononenko,Wei,Duda} were employed for the ranking of discriminative features. ReliefF Algorithm \cite{Kononenko} selects the features based on the difference of the weights assigned to a pair of features in the neighborhood.
 Higher ranks are assigned to the features with different values for different classes. In Feature based Neighborhood Component Analysis (fNCA)\cite{Wei}, feature weights are allocated in a way that an objective function that measures the average leave-one-out regression loss over the training data is minimized. Fisher Score \cite{Duda} assigns a score to every feature by measuring the ratio of variability between different and similar features of the training data sets .

 \begin{table}[!tbp]
\centering
\caption{Distribution of types and sizes for segmented nodules}
\centering
\begin{tabular}{|c|c|c|c|}
\hline
\begin{tabular}[c]{@{}c@{}}Type of \\ \\ nodule\end{tabular} & \begin{tabular}[c]{@{}c@{}}No. of nodules\\ (training cohort)\end{tabular} & \begin{tabular}[c]{@{}c@{}}No. of nodules\\ (test cohort)\end{tabular} & \begin{tabular}[c]{@{}c@{}}Min.-Max.\\ Dia. of nodules \\(mm)\end{tabular} \\ \hline
Malignant                                                    & 165                                                                        & 50                                                                     & 11.39-133.24                                                                     \\ \hline
Benign                                                       & 35                                                                         & 30                                                                     & 4.44- 83.25                                                                      \\ \hline
\end{tabular}
\end{table}
 
 \par Additionally, a selection of the three competent unsupervised feature selection algorithms reported in the literature was also made for the new features ranking. The algorithm in \cite{Hall} selects the features exhibiting minimum correlation with each other, whereas the Laplacian score \cite{He} computes a score for each feature to reflect its locality preserving power. Multi-cluster feature selection (MCFS) \cite{Cai} technique selects and ranks the features by measuring the correlations between different features by solving the process as a sparse Eigen-problem and a L1- regularized least squares problem.
 
 \begin{figure}[!tbp]
\centering
{\includegraphics[width=3in,height=5in]{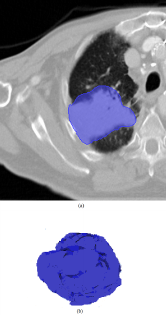}}
\caption{Segmentation of a malignant nodule from Lung1\cite{Hugo} database in (a) 2-D axial slice  (b) 3-D axial slice}
\label{fig}
\end{figure}

While analyzing the ranking scores assigned by the 3 chosen supervised ranking algorithms, it was observed that scores of each of the 52 features varied greatly and were not synchronized. This trend can be observed in Table 2 in the individual scores assigned by Fisher Score, ReliefF network and fNCA to the top 20 features. For example, Fisher score ranked Surface Volume ratio(SVR) at 18, while ReliefF network gave SVR the 4th rank and fNCA assigned it the 2nd rank. The vast difference in designated scores by the 3 renowned algorithms indicate that feature ranking methods can heavily influence the classification using machine learning methods and the results will differ with each of the 3 selection methods. Since there is a lack of discussion on the performance of feature ranking algorithms for machine learning classifiers, the published classification studies have simply employed any one of the popular ranking methods. We have already shown that by changing the algorithm, the classifier performance differs and this leaves a room of debate on the reported nodule classification performance in the literature. 
Similar variation in the assigned scores was observed for the 3 unsupervised ranking algorithms namely Laplacian score, Similarity based on minimum correlation and MCFS. Features ranked highly by one method gets lower ranks by the others and vice versa. In order to overcome the discrepancies in the ranking of every feature, each ranking algorithm was given equal weight-age and the 3 ranking scores from supervised and unsupervised ranking methods were averaged separately to obtain 2 lists of final feature ranking. The top 20 ranked features out of 51 discriminative features according to their scores from both the group of algorithms are shown in Table 2 and Table 3.
 
\begin{figure}[!tbp]
\centerline{\includegraphics[width=3.5in,height=3in]{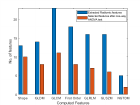}}
\caption{Feature extraction and Feature reduction with respect to feature class}
\label{fig}
\end{figure}
\par Out of the 20 features ranked, only top 2 features from both the lists are defined in the paper to utilize the paper space efficiently. The top 2 discriminative features obtained from the supervised ranking approach were Surface Volume Ratio and Small Dependence High Gray Level Emphasis (SDHGLE). 

The first top ranked feature, Surface Volume ratio (SVR) is defined as :
\begin{equation}
  SVR=\frac{S}{V}  
\end{equation}
Here $S$ is the surface area and $V$ denotes the volume of the sample nodule. 

Second highest ranked feature, Small Dependence High Gray Level Emphasis(SDHGLE) computes the collective distribution of high gray levels having minimum dependence and is defined as follows \cite{Weszka}:
\begin{equation}
 SDHGLE = \frac{\sum_{x=1}^{N_{i}} \sum_{y=1}^{N_{dz}}\frac{p(x,y){x^{2}}} {y^{2}{ }}}{\sum_{x=1}^{N_{i}} \sum_{y=1}^{N_{dz}}P(x,y)}
\end{equation} 
Here $P(x,y)$ is the dependence matrix and $p(x,y)$ is the normalized dependence matrix. $N_{i}$ is the number of distinct gray-values in the nodule and 
$N_{dz}$ are the dependent zones in the nodule.

Using the supervised ranking approach, the top 2 discriminative features towards lung cancer are Small Dependence Low Gray Level Emphasis and  Zone Variance. 
Small Dependence Low Gray Level Emphasis(SDLGE) describes the  collective distribution of small dependence with smaller values of gray-level and is defined as \cite{Orozco}:

\begin{equation}
   SDLGE = \frac{\sum_{x=1}^{N_{i}} \sum_{y=1}^{N_{dz}}\frac{p(x,y)} {y^{2}{x^{2} }}}{\sum_{x=1}^{N_{i}} \sum_{y=1}^{N_{dz}}P(x,y)}
\end{equation}
Here $P(x,y)$ is the dependence matrix and $p(x,y)$ is the normalized dependence matrix. $N_{i}$ is the number of distinct gray values in the nodule and $N_{dz}$ are the dependent zones in the nodule. 

The second highest ranked feature obtained from unsupervised ranking approach is zone variance(ZV). It measures the variability in zone size volumes and is given as follows \cite{Zwanenburg}:

\begin{equation}
ZV=\sum_{x=1}^{N_{i}} \sum_{y=1}^{N_{dz}}P(x,y){(y- \mu)}^{2}
\end{equation}

Here \begin{equation}
    \mu=\sum_{x=1}^{N_{i}} \sum_{y=1}^{N_{dz}}P(x,y)y
\end{equation}
\subsection{Training the nodule classifier}

To perform the nodule classification, Support Vector Machine(SVM)\cite{Orozco} and Naive Bayes\cite{Shi} were chosen as the supervised machine learning models to perform nodule classification. The SVM is known for its good performance towards complex classification problems and Naive Bayes has been used as a second best choice to SVM for classification.

The first 2, 4, 8, 12, 16 features and then all the top 20 ranked features were selected from the features ranking lists of both selection algorithms to train  the SVMs and Naive Bayes classifiers respectively. Finally, the two classifiers were cross-validated using 10-fold cross-validation. In total, 12 SVMs and 12 Naive Bayes classifiers were trained using 6 combination of features from supervised and unsupervised feature selection methods.
\section{Results}

The training and testing of SVMs and Naive Bayes classifiers were performed using MATLAB 2018b platform. The programming of supervised as well as unsupervised ranking algorithms were also done in MATLAB environment. 
A total of 50 malignant and 30 benign test nodules were classified using the 12 SVMs and 12 Naive Bayes classifiers.

The nodule classification results were evaluated using the performance metric of accuracy, specificity and sensitivity which are defined as follows \cite{Thibault}:
\begin{equation}
Accuracy =\frac{t_p+t_n}{(t_p+f_n+f_p+t_n)}
\end{equation}

\begin{equation}
Specificity =\frac{tn}{(tp+fp)}   
\end{equation}
\begin{equation}
Sensitivity =\frac{tn}{(tn+fn)}   
\end{equation}

Here $t_p$ stands for true positive and denotes the number of cancer patients correctly diagnosed. The $t_n$ term is true negative and describes the number of benign nodules correctly diagnosed as benign. $f_p$ is false positive and represents the number of cases wrongly diagnosed with cancer and $f_n$ is false negative and describes the cancerous nodules mis-classified as benign.
The classification performance of Naive Bayes classifier using 2, 4, 8, 12, 16 and 20 selected features from supervised and unsupervised ranking methods is shown in Figure 4. From supervised feature selection, Naive Bayes gave the best diagnosis using 2 features and classified 50 malignant and 27 benign nodules correctly with an accuracy of 97.468\%, sensitivity of 100\% and specificity of 93.1\% respectively. Using unsupervised approach, the classifier gave the best performance using 2 features and diagnosed 23 malignant and 29 benign nodules. The accuracy was 65.82\%, sensitivity was 46\% and specificity was 100\%.

Based on the computed performance metric, the performance of SVM classifier with 2, 4, 8, 12 , 16 and 20 radiomic features is demonstrated in Figure 5 for supervised as well as for unsupervised ranking approaches. From the inspection of classification results using supervised ranking method, the best classification result was achieved with 8 and 12 features with an accuracy of 100\%, sensitivity of 100\% and specificity of 100\% respectively. From the unsupervised ranking approach, the 2 feature SVM classifier showed the highest performance by classifying 40 malignant and 29 benign nodules with an accuracy of 87.34\%, sensitivity of 79.75\% and specificity of 100\% respectively.

\begin{table*}[!tbp]
\caption{Top 20 ranked features according to the supervised ranking algorithms}
\label{tab:my-table}
\begin{tabular}{|l|l|l|l|l|l|}
\hline
\textbf{\begin{tabular}[c]{@{}l@{}}Feature\\  rank\end{tabular}} & \textbf{Radiomic Feature}                                                               & \multicolumn{1}{c|}{\textbf{Fisher Score}} & \textbf{\begin{tabular}[c]{@{}l@{}}ReliefF\\  net. score\end{tabular}} & \textbf{\begin{tabular}[c]{@{}l@{}}fNCA\\ score\end{tabular}} & \textbf{\begin{tabular}[c]{@{}l@{}}Avg. ranking\\ Score\end{tabular}} \\ \hline
1                                                                & Surface Volume Ratio                                                                    & 18                                         & 4                                                                         & 2                                                            & 8                                                                     \\ \hline
2                                                                & \begin{tabular}[c]{@{}l@{}}Small Dependence High \\ \\ Gray Level Emphasis\end{tabular} & 12                                         & 2                                                                         & 13                                                           & 11                                                                    \\ \hline
3                                                                & Difference Average                                                                      & 13                                         & 11                                                                        & 11                                                           & 12                                                                    \\ \hline
4                                                                & Joint Entropy                                                                           & 4                                          & 13                                                                        & 19                                                           & 14.33333                                                              \\ \hline
5                                                                & 90Percentile                                                                            & 27                                         & 3                                                                         & 44                                                           & 14.66667                                                              \\ \hline
6                                                                & Sum Entropy                                                                             & 7                                          & 29                                                                        & 30                                                           & 16.33333                                                              \\ \hline
7                                                                & Interquartile Range                                                                     & 15                                         & 5                                                                         & 26                                                           & 17.33333                                                              \\ \hline
8                                                                & Sum Squares                                                                              & 22                                         & 7                                                                         & 34                                                           & 17.66667                                                              \\ \hline
9                                                                & Variance                                                                                & 23                                         & 10                                                                        & 25                                                           & 17.66667                                                              \\ \hline
10                                                               & Dependence Variance                                                                      & 2                                          & 42                                                                        & 40                                                           & 18.33333                                                              \\ \hline
11                                                               & Maximum3DDiameter                                                                       & 17                                         & 34                                                                        & 24                                                           & 18.33333                                                              \\ \hline
12                                                               & Least Axis                                                                              & 3                                          & 39                                                                        & 15                                                           & 19                                                                    \\ \hline
13                                                               & \begin{tabular}[c]{@{}l@{}}High GrayLevel \\ Emphasis\end{tabular}                      & 21                                         & 8                                                                         & 3                                                            & 19.33333                                                              \\ \hline
14                                                               & Run Percentage                                                                          & 1                                          & 32                                                                        & 17                                                           & 19.33333                                                              \\ \hline
15                                                               & Inverse Variance                                                                        & 10                                         & 16                                                                        & 31                                                           & 20                                                                    \\ \hline
16                                                               & Major Axis                                                                              & 14                                         & 46                                                                        & 21                                                           & 21                                                                    \\ \hline
17                                                               & \begin{tabular}[c]{@{}l@{}}Maximum2DDiameter\\ Row\end{tabular}                         & 9                                          & 44                                                                        & 33                                                           & 21                                                                    \\ \hline
18                                                               & \begin{tabular}[c]{@{}l@{}}Large Area High\\ GrayLevel Emphasis\end{tabular}            & 21                                         & 21                                                                        & 25                                                           & 21                                                                    \\ \hline
19                                                               & Cluster Prominence                                                                      & 38                                         & 6                                                                         & 9                                                            & 21.66667                                                              \\ \hline
20                                                               & Surface Area                                                                            & 11                                         & 48                                                                        & 32                                                           & 22                                                                    \\ \hline
\end{tabular}
\end{table*}

\begin{table*}[!tbp]
\caption{Top 20 ranked features according to the unsupervised ranking algorithm}
\label{tab:my-table}
\begin{tabular}{|l|l|l|l|l|l|}
\hline
\begin{tabular}[c]{@{}l@{}}Feature\\ rank\end{tabular} & \begin{tabular}[c]{@{}l@{}}Radiomic  Feature\end{tabular} & \begin{tabular}[c]{@{}l@{}}Laplacian \\ Score\end{tabular} & \begin{tabular}[c]{@{}l@{}}Minimum \\ Corr.Score\end{tabular} & MCFS & \begin{tabular}[c]{@{}l@{}}Avg.\\ score\end{tabular} \\ \hline
1 & \begin{tabular}[c]{@{}l@{}}Small Dependence Low\\ Gray Level Emphasis\end{tabular} & 40 & 3 & 3 & 15.33333 \\ \hline
2 & Zone variance & 3 & 2 & 43 & 16 \\ \hline
3 & Uniformity & 29 & 1 & 19 & 16.33333 \\ \hline
4 & LargeAreaEmphasis & 2 & 8 & 44 & 18 \\ \hline
5 & MaximumProbability & 46 & 4 & 4 & 18 \\ \hline
6 & LowGrayLevelEmphasis & 42 & 7 & 7 & 18.66667 \\ \hline
7 & LeastAxis & 28 & 9 & 23 & 20 \\ \hline
8 & Correlation & 33 & 14 & 14 & 20.33333 \\ \hline
9 & LargeAreaLowGrayLevelEmphasis & 7 & 6 & 48 & 20.33333 \\ \hline
10 & RunLengthNonUniformityNormalized & 41 & 16 & 6 & 21 \\ \hline
11 & LargeAreaHighGrayLevelEmphasis & 21 & 21 & 21 & 21 \\ \hline
12 & GrayLevelNonUniformityNormalized & 52 & 11 & 2 & 21.66667 \\ \hline
13 & Maximum2DDiameterColumn & 25 & 10 & 32 & 22.33333 \\ \hline
14 & ClusterProminence & 4 & 12 & 51 & 22.33333 \\ \hline
15 & LowGrayLevelRunEmphasis & 51 & 13 & 5 & 23 \\ \hline
16 & Maximum2DDiameterSlice & 21 & 23 & 27 & 23.66667 \\ \hline
17 & DependenceNonUniformityNormalized & 44 & 19 & 9 & 24 \\ \hline
18 & LongRunLowGrayLevelEmphasis & 38 & 17 & 17 & 24 \\ \hline
19 & HighGrayLevelEmphasis & 12 & 22 & 38 & 24 \\ \hline
20 & ShortRunLowGrayLevelEmphasis & 47 & 18 & 12 & 25.66667 \\ \hline
\end{tabular}
\end{table*}

\section{Discussion}
Evidently, the results achieved in terms of accuracy, specificity and sensitivity using supervised feature ranking techniques have led to superior nodule classification. This is true for SVMs as well as Naive Bayes as demonstrated through Figure 4 and Figure 5. The minimum accuracy achieved with supervised ranking approach is 77.21\% from SVM classifier while with unsupervised feature selection, it is 43.08\% using Naive Bayes . 

Between the two classifiers, the classification results have proven SVM to be a better choice for nodule classification with an accuracy, sensitivity and specificity all of 100\% using 8 as well as 12 supervised features training respectively. Moreover, the quantitative SVM classification results for the remaining 2, 4, 16 and 20 features are also overall better than the Naive Bayes classifier(supervised and unsupervised feature selection).

When we evaluated the optimal number of features from the performance metric used for classification, we found out that overall 2 features trained classifiers produced excellent classification results.Although SVM with 8 features using supervised ranking is the only case with a slightly larger number of features demonstrating good performance , SVM with 2 features using supervised ranking is a second close. The accuracy of the latter SVM is 98.8\%, sensitivity is 98\% and specificity is 100\% respectively. This clearly shows that even in this case, 2 feature SVM using supervised ranking can be considered an alternative of the 8 feature SVM using supervised ranking. 

\par Keeping in view the aforesaid argument, it is concluded that 2 feature trained classifiers using supervised ranking approach have proven to be the best option for lung nodule classification using machine learning classifiers. This pattern also clearly entails that nodule classification performance may not necessarily improve by increasing the number of radiomic features for nodule classifier using unsupervised or unsupervised ranking approach. 

\par Finally, it is safe to claim that that supervised ranking methods generally perform better overall than the unsupervised ranking algorithms for feature selection and can become a preferred approach for nodule classification.

\section{Conclusion}

In this research work, the impact of various feature selection techniques and number of features is explored on machine learning classifiers. This is an important study because correct number of features and feature selection method can result in improved cancer detection using neural networks and machine learning classifiers. 

The study has revealed that supervised feature selection techniques accounts for superior classification results.
The research experiments also showed that increasing the number of radiomic features does not show any significant improvement in the classification results whether the selection is done using unsupervised or unsupervised feature selection methods. Therefore very good classification can be performed with a fewer features trained classifier. Hence, the study contributes some important findings in radiomic features selection for the future radiomic based nodule classifiers. The presented analysis can be taken up for further exploratory studies using other machine learning methods and by increasing the number of radiomic features from 20 upward for nodule classification.

\section*{Acknowledgment}
The authors would like to acknowledge all the contributors who made the lung CT data sets publicly available on TCIA site \\ (http://www.cancerimagingarchive.net/).

\begin{figure*}[!tbp]
\centering
\begin{subfigure}[b]{0.8\textwidth}
   \includegraphics[width=1\linewidth]{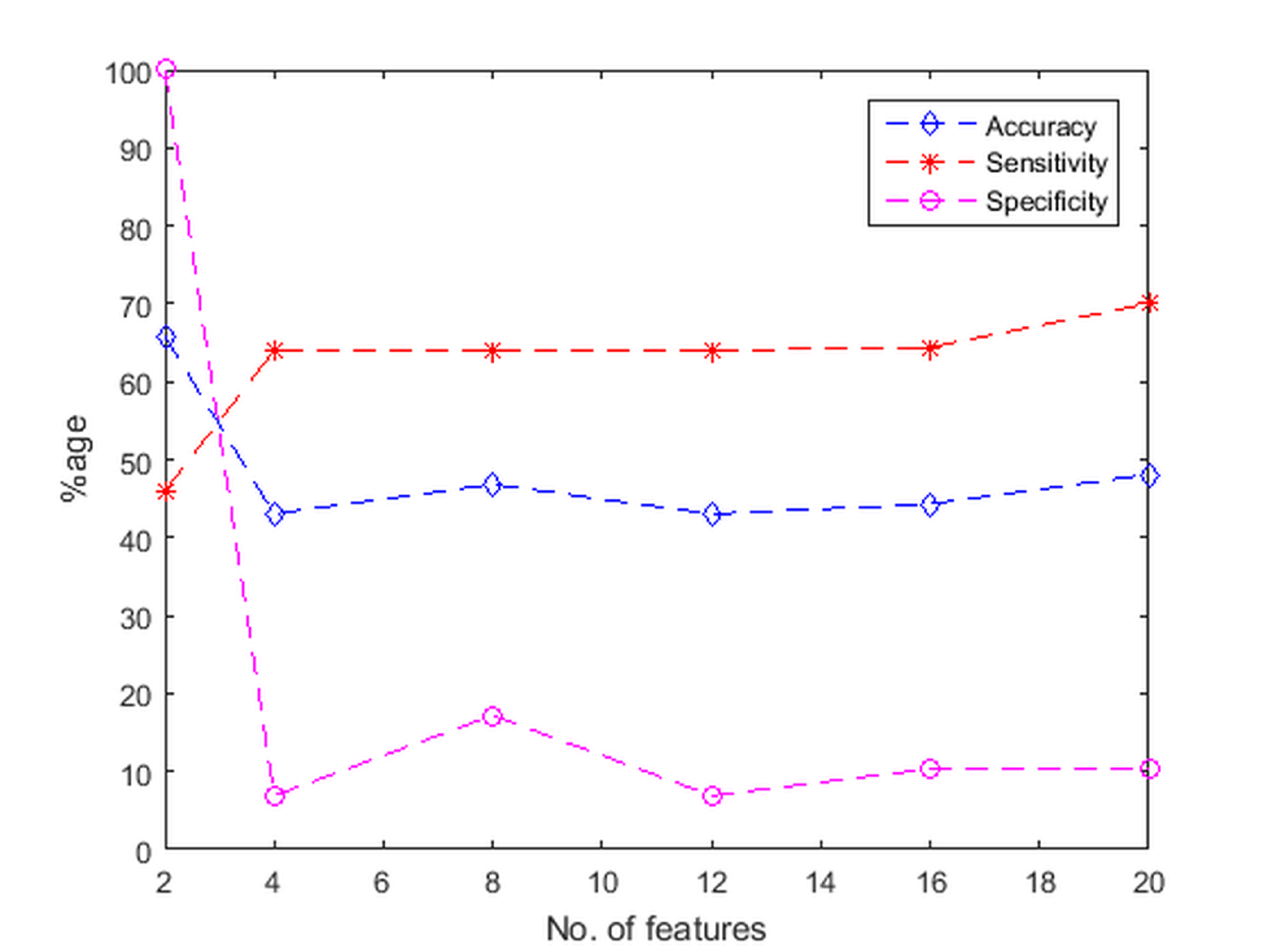}
   \caption{}
   \label{fig:Ng1} 
\end{subfigure}
\begin{subfigure}[b]{0.8\textwidth}
   \includegraphics[width=1\linewidth]{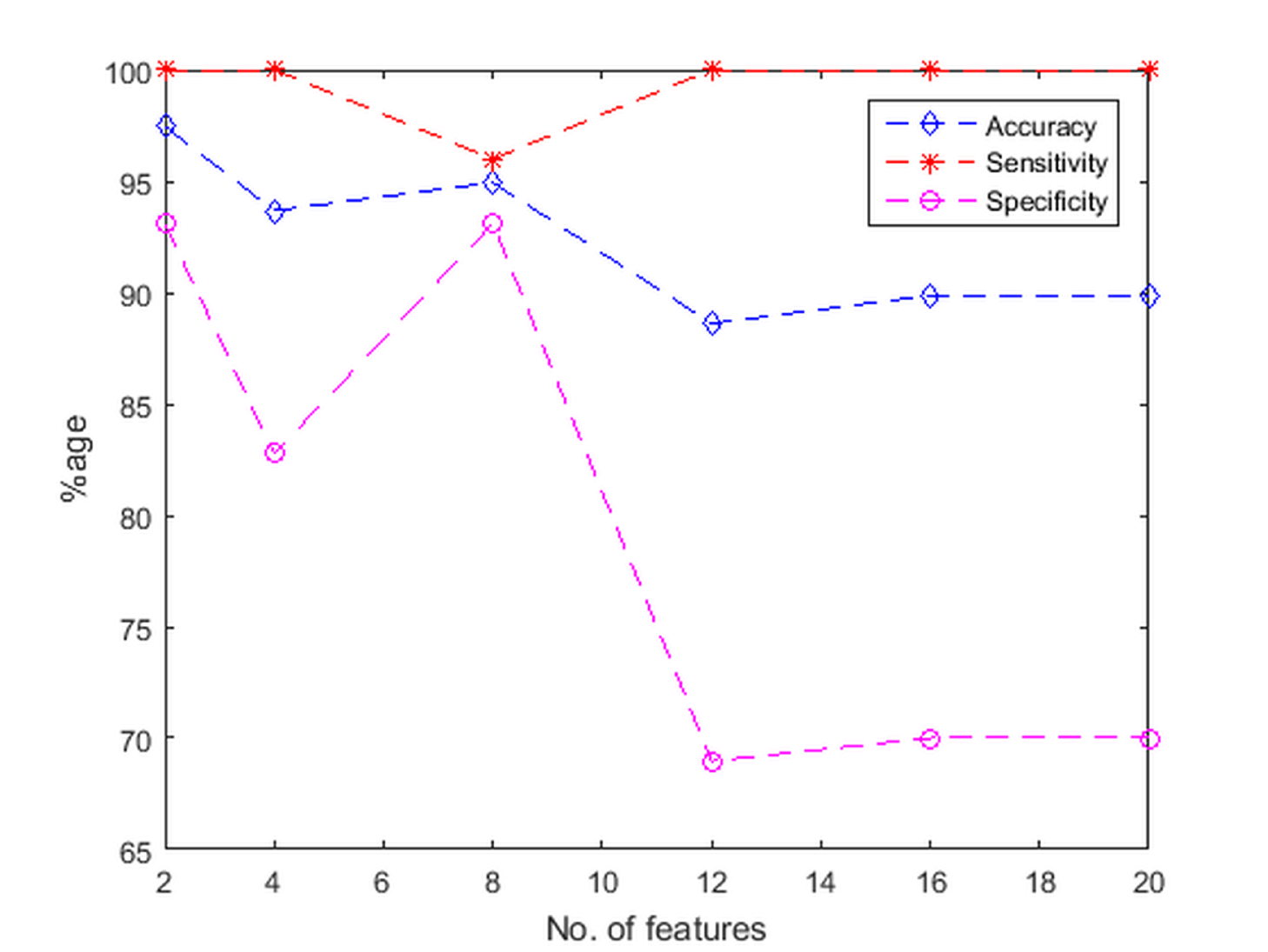}
   \caption{}
   \label{fig:Ng2}
\end{subfigure}
   \caption{Naive Bayes Classification performance (a)using unsupervised feature ranking (b) using supervised feature ranking}\label{bs1}
   \centering
\end{figure*}

\begin{figure*}[!tbp]
\centering
\
\begin{subfigure}[b]{0.8\textwidth}
   \includegraphics[width=1\linewidth]{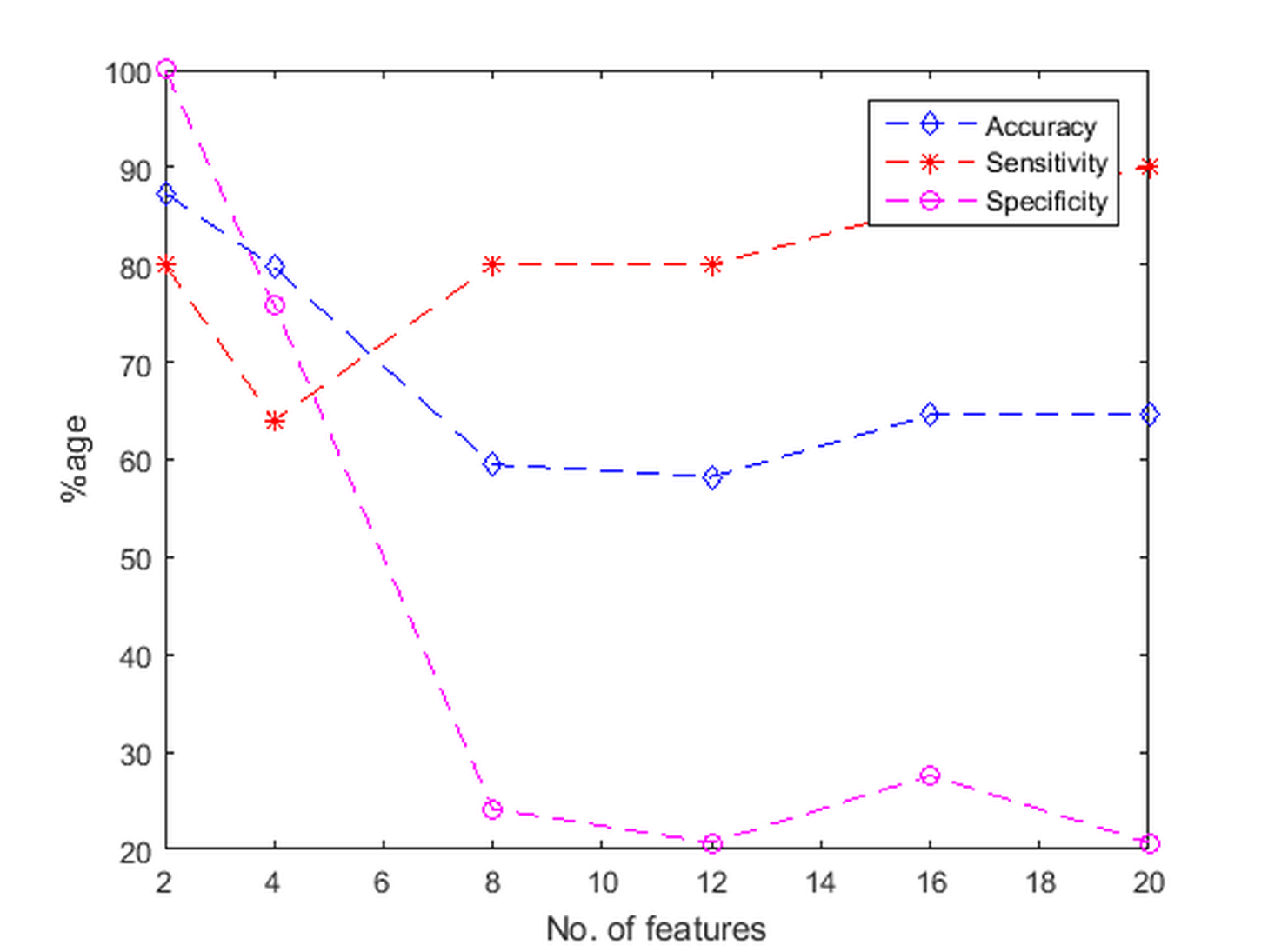}
   \caption{}
   \label{fig:Ng1} 
\end{subfigure}
\begin{subfigure}[b]{0.8\textwidth}
   \includegraphics[width=1\linewidth]{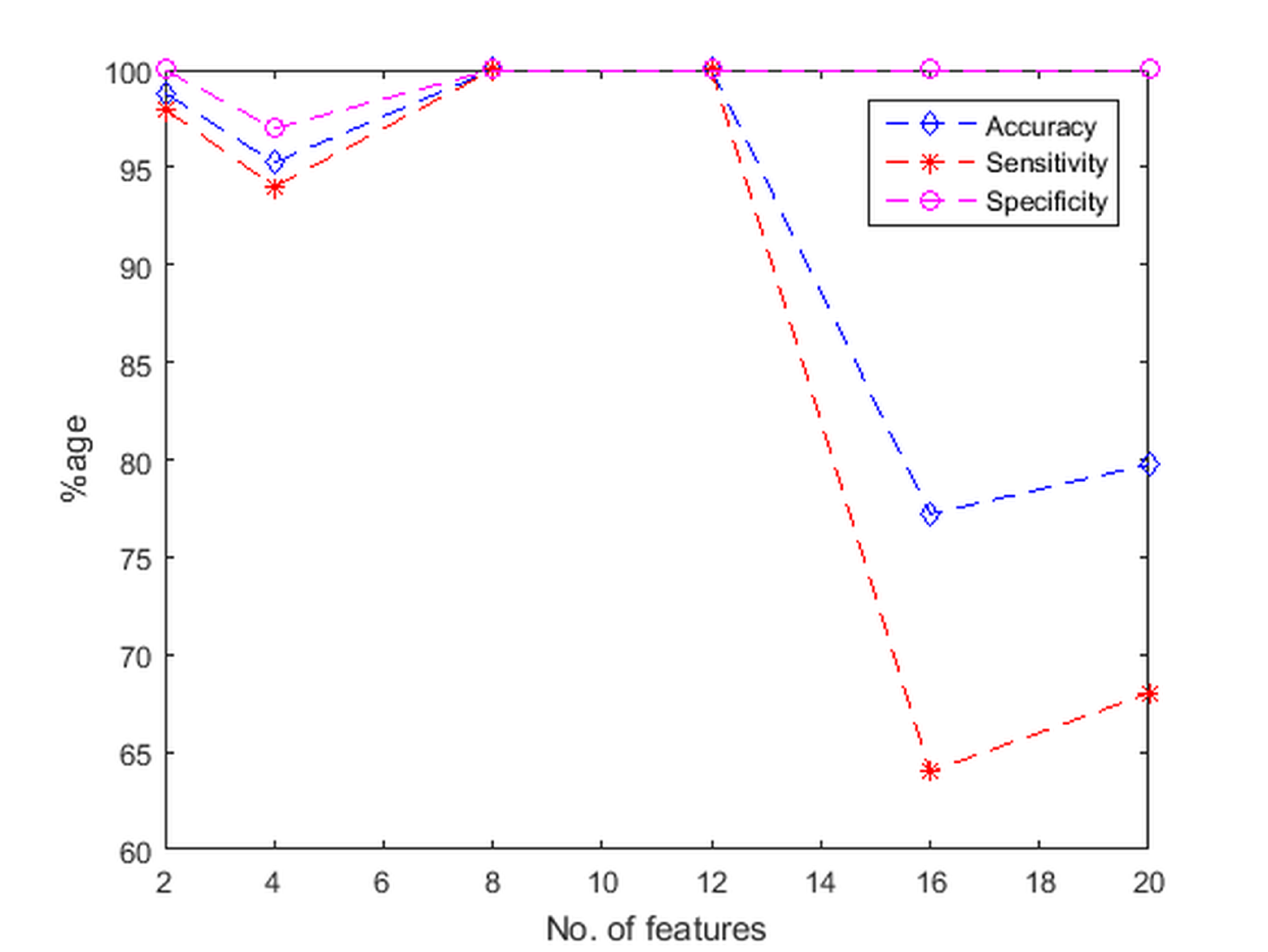}
   \caption{}
   \label{fig:Ng2}
\end{subfigure}  
       \centering
   \caption{SVM Classification performance (a) using unsupervised feature ranking  (b)using supervised feature ranking }\label{bs1}
   \centering
\end{figure*}



\begin{thebibliography}{40}
%
%
\bibitem{Siegel} R. L. Siegel , K. D.  Miller, and A.  Jemal, “A. Cancer statistics,”  CA: A Cancer J. for Clin., 68, 7–30,2018, DOI:10.3322/caac.21442.

\bibitem{Ayman}Ayman El-Baz, Garth M. Beache, Georgy Gimel'farb, et al., “Computer-Aided Diagnosis Systems for Lung Cancer: Challenges and Methodologies,” International Journal of Biomedical Imaging, vol., 2013

\bibitem{Gillies}R. J. Gillies, P. E.  Kinahan, and  H. Hricak, “Radiomics: images are more than pictures, they are data,” Radiology 278, 563–577, 2016.

\bibitem{Chen}C. Chen et al., “Radiomic features analysis in computed tomography images of lung nodule classification,”  PLoS ONE 13, DOI: https://doi.org/10.1371/journal.pone.0192002 ,2018.


\bibitem{Jingchen} M. Jingchen, W.Qian, R. Yacheng, H. Haibo, and Z. Jun, “Automatic lung nodule classification with radiomics
Approach,” In Proceedings of SPIE, vol. 9789, DOI: 10.1117/12.2220768, SPIE, 2016.

\bibitem{Tu} S. J. Tu, C. W. Wang, K. T. Pan , Y. C. Wu , and C. T. Wu, “Localized thin -section  CT with radiomics feature extraction and machine learning to classify early detected pulmonary nodules from lung cancer screening,”  Phys Med Biol.,2018.

\bibitem{Kadir}T. Kadir , F. Gleeson, “ Lung cancer prediction using machine learning and advanced imaging techniques,” Transl. Lung Cancer Res;7:304-12,2018.
 
\bibitem{Choi} W. Choi et al. , “Radiomics analysis of pulmonary nodules in low-dose ct for early detection of lung cancer,” Med.Phys. 45, 1537–1549 ,2018.


\bibitem{Wu}W. Wu. et al., “Exploratory study to identify radiomics classifiers for lung cancer histology,” Front. Oncol. 6, 71, 2016, DOI: 10.3389/fonc.2016.00071.

\bibitem{Sun}T. Sun, R. Zhang, J. Wang, X. Li, X. Guo , “Computer-Aided Diagnosis for Early-Stage Lung Cancer Based on Longitudinal and Balanced Data,” PLoS ONE 8(5): e63559,2-13,2013, https://doi.org/10.1371/journal.pone.
0063559

\bibitem{Shi}J H. Shi, Y. Liu, Naïve Bayes vs. Support "Vector Machine: Resilience to Missing Data", Berlin, Heidelberg:Springer Berlin Heidelberg, pp. 680-687, 2011.

\bibitem{Shakir1}Hina Shakir and Yiming Deng and Haroon Rasheed and Tariq Mairaj Rasool Khan, "Radiomics based likelihood functions for cancer diagnosis",Scientific Reports,9, 2019, doi: https://doi.org/10.1371/journal.pone.0200721.
 


\bibitem{Hugo} A. Hugo et al., “ Data from nsclc-radiomics. the cancer imaging archive,”2015, DOI: http://doi.org/10.7937/K9/TCIA.2015.
PF0M9REI.



\bibitem{Armato}S. Armato et al., “ Data from lidc-idri. the cancer imaging archive,”,2015, DOI: http://doi.org/10.7937/K9/TCIA.2015.
LO9QL9SX .
\bibitem{Armato1}S. Armato et al., “Spie-aapm-nci lung nodule classification challenge dataset. the cancer imaging archive,”2015, DOI: http://doi.org/10.7937/K9/TCIA.2015.UZLSU3FL .

 
\bibitem{Fedorov}A. Fedorov et al., “3d slicer as an image computing platform for the quantitative imaging network,” Magn. Reson.I maging 30, 1323 – 1341, 2012. Quantitative Imaging in Cancer.

\bibitem{Velazquez} V. Emmanuel et al., “Volumetric ct-based segmentation of nsclc using 3d-slicer,” IEEE Transactions on Biomed. Eng. 3, 2013, DOI: 10.1038/srep03529 .

\bibitem{Griethuysene} J. J. van Griethuysen et al., “Computational radiomics system to decode the radiographic phenotype,” Cancer Res. 77, 2017,e104–e107, DOI: 10.1158/0008-5472.CAN-17-0339.


\bibitem{Kononenko} I. Kononenko, E. Šimec  and M. Robnik-Šikonja, “Overcoming the myopia of inductive learning algorithms with relieff,”  Appl. Intell. 7, 39–55, 1997, DOI: 10.1023/A:1008280620621.

\bibitem{Wei} Y. Wei, W. Kuanquan  and  Z. Wangmeng, “ Neighborhood component feature selection for high-dimensional data,” J. Comput. 7, 161–168, 2012, DOI: 10.4304/jcp.7.1.161-168.

\bibitem{Duda}R. Duda, P. Hart and D. G. Stork, “ Pattern Classification,” JOHN WILEY SONS, 2001.


\bibitem{Hall} M. A. Hall,"Correlation-based feature selection for machine learning," Tech. Rep. ,1999.


\bibitem{He} X. He, D. Cai, and P. Niyogi ,”Laplacian score for feature selection,” In Proceedings of the 18th International Conference on Neural Information Processing Systems, NIPS’05, 507–514 ,2005.


\bibitem{Cai}D. Cai, C. Zhang and X. He, “Unsupervised feature selection for multi-cluster data,”  In Proceedings of the 16th ACM SIGKDD International Conference on Knowledge Discovery and Data Mining, KDD ’10, 333–342,  2010.
\bibitem{Weszka}J. S. Weszka, C.R. Dyer,  A. Rosenfeld, “A comparative study of texture measures for terrain classification,” IEEE Transactions on Systems, Man, and Cybernetics , 4(SMC-6):269–285,1976.
\bibitem{Orozco}H. M. Orozco, O.O.V. Villegas, V.G.C. Sánchez, H.D.J.O. Domínguez, and M.D.J.N. Alfaro, Automated system for lung nodules classification based on wavelet feature descriptor and support vector machine, Biomed Eng Online 14, 2015.
\bibitem{Zwanenburg}A. Zwanenburg , S. Leger, M.  Vallières, and S. Löck, “Image biomarker standardisation initiative - feature definitions,” In eprint arXiv: 1612. 07003, 2016
\bibitem{Haarlick}R. M. Haarlick, “Statistical and structural approaches to texture,” Proceedings of the IEEE ,7(5):786–804, 1979
\bibitem{Thibault}G. Thibault, B.  Fertil, C. Navarro, S. Pereira, P. Cau, N. Levy, J.
Sequeira, J.  Mari, “Texture Indexes and Gray Level Size Zone Matrix. Application to Cell
Nuclei Classification,” Pattern Recognition and Information Processing (PRIP), 140-145, 2009.
\end{thebibliography}
\end{document}